\documentclass[runningheads,dvipsnames]{llncs}

\usepackage{graphicx}
\usepackage{epsfig}
\usepackage{amsmath}
\usepackage{amssymb}
\usepackage[dvipsnames]{xcolor}
\usepackage{enumerate} 
\usepackage{amssymb}
\usepackage{pifont} 
\newcommand{\cmark}{\ding{51}}
\newcommand{\xmark}{\ding{55}}
\usepackage{tabularx}
\usepackage{booktabs}
\usepackage[hidelinks]{hyperref}

\newcolumntype{h}{>{\hsize=1.1\hsize}X}
\newcolumntype{b}{>{\hsize=.45\hsize}X}
\newcolumntype{s}{>{\hsize=.14\hsize}X}

\begin{document}

\title{Evidence-Driven Differential Diagnosis\\ of Malignant Melanoma}
\titlerunning{Evidence-Driven Differential Diagnosis of Malignant Melanoma}

\author{Naren Akash R J \and Anirudh Kaushik \and
Jayanthi Sivaswamy}
\index{Akash R J, Naren \and Kaushik, Anirudh \and Sivaswamy, Jayanthi}
\authorrunning{Akash and Kaushik et al.}
\institute{Center for Visual Information Technology\\ International Institute of Information Technology Hyderabad, India\\
\href{https://cvit.iiit.ac.in/mip/projects/meldd}{\email{https://cvit.iiit.ac.in/mip/projects/meldd}}}

\maketitle 
\begin{abstract} 

We present a modular and multi-level framework for the differential diagnosis of malignant melanoma. Our framework integrates contextual information and evidence at the lesion, patient, and population levels, enabling decision-making at each level. We introduce an anatomic-site aware masked transformer, which effectively models the patient context by considering all lesions in a patient, which can be variable in count, and their site of incidence. Additionally, we incorporate patient metadata via learnable demographics embeddings to capture population statistics. Through extensive experiments, we explore the influence of specific information on the decision-making process and examine the tradeoff in metrics when considering different types of information. Validation results using the SIIM-ISIC 2020 dataset indicate including the lesion context with location and metadata improves specificity by $17.15\%$ and  $7.14\%$, respectively, while enhancing balanced accuracy. The code is available at \href{https://github.com/narenakash/meldd}{\email{https://github.com/narenakash/meldd}}.

\end{abstract}
\section{Introduction}

Melanoma is the most invasive form of skin cancer with the highest mortality rate; its incidence is rising faster among other types of cancer and is projected to increase by $57\%$ globally by 2040, leading to an estimated $68\%$ rise in mortality \cite{jama_caseburden}. When caught early, it has an increased survival rate and tends to have a better prognosis. However, melanoma is a complex and heterogeneous disease which makes accurate early recognition non-trivial and challenging. Melanoma can masquerade/appear as benign lesions and benign pigmented lesions can resemble melanoma, making diagnosis difficult even for skilled dermatologists \cite{acadderm_misdiagnosis}.

A dermatologist's expertise to discriminate between benign moles/nevi and melanoma relies on the recognition of morphological features through the ABCD criteria \cite{acadderm_abcd}, applying the 7-point checklist \cite{derm_7ptchecklist}, overall pattern recognition and differential recognition of the ugly duckling nevi \cite{derm_uglyduckling}. Most nevi in a patient tend to be similar and can be grouped into a few clusters based on morphological similarity \cite{investderm_clustering}. Any nevus that deviates from a consistent pattern within an individual is an outlier or an ugly duckling which is taken to be a suspicious lesion \cite{jama_uglyduckling}. Dermatologists utilize an intra-patient lesion-focused as well as comparative analysis, recognising overall patterns to identify ugly ducklings before forming a provisional diagnosis \cite{derm_melprocess}. This approach considers the characteristics of individual lesions while also taking into account the context of the patient's overall nevi distribution, leading to improved accuracy in identifying melanoma \cite{derm_abcdef}. Furthermore, patient demographics, including age, sex, and anatomical site, are risk factors to consider in the differential diagnosis of melanoma \cite{envres_demographics}. Age-related susceptibility, anatomical site variations, and sex-specific characteristics contribute to the complexity of melanoma diagnosis. 

Recent advances in deep learning techniques have led to an interest in the development of AI models for dermatology. The integration of AI systems into clinical workflows has the potential to improve the speed and accuracy of melanoma diagnosis, saving lives. Existing deep learning methods have reported good diagnostic accuracy in the classification of skin lesions. These use largely lesion-focused approaches and include the seven-point checklist \cite{jbhi_derm7pt}, hierarchical structures \cite{pr_taxonomies}, lesion segmentation \cite{jbhi_dermaknet}, and ABCD-based medical representations \cite{cvpr_abcd}. Despite integrating clinical knowledge, most existing methods have not fully harnessed the potential of the clinician's comprehensive diagnostic process and strategy. While some approaches, such as CI-Net \cite{tmi_cinet}, incorporate zoom-observe-compare processes, they focus only on individual lesion characteristics. The UDTR framework \cite{miccai_udtr} incorporates contextual information of lesions to model ugly ducklings but assumes a fixed number of contextual lesions. No attempt has been made by any approach so far to take into account a richer set of information that clinicians rely on for melanoma diagnosis \cite{investderm_complexity}. These include lesion counts in a patient, which can be variable, lesion location in the body and patient demographic information. Further, existing models are essentially a black box, with no or limited explainability that too post facto on the basis of visualisation of activations and so on. 

We wish to design a melanoma recognition solution that incorporates a rich set of information similar to the clinical practice. Our aim is to understand how the addition of specific information influences the decision-making process. An understanding of the sensitivity-specificity tradeoff when considering different types of information can make a method more transparent. This transparency can enable clinicians to critically evaluate the AI system's recommendations and ensure that decisions align with their clinical expertise and patient-specific circumstances. In this paper, we present a method for melanoma recognition with the following contributions:
\begin{itemize}
    \item A modular, multi-level framework for evidence-based differential diagnosis of melanoma. This offers a solution to holistically integrate evidence at multiple levels (lesion, patient and population).
    \item A solution based on a masked transformer to utilize \textit{variable-count} context lesions from a patient along with their anatomic location and metadata such as age and sex.
    \item Insights on the role of various information in melanoma recognition, based on validation results of the proposed approach on the 2020 SIIM-ISIC dataset.
\end{itemize}

\section{Method}

In our design, features are first extracted for each given lesion image (using a CNN) and are grouped patient-wise. The \textit{context} of lesions within each patient is captured using a transformer encoder with masked self-attention. Age, sex, and anatomical site embeddings are included as \textit{supporting evidence}, which along with lesion context, are fed into the classification layers to predict melanoma.\\

\begin{figure}[t]
    \centering
    \includegraphics[width=\textwidth]{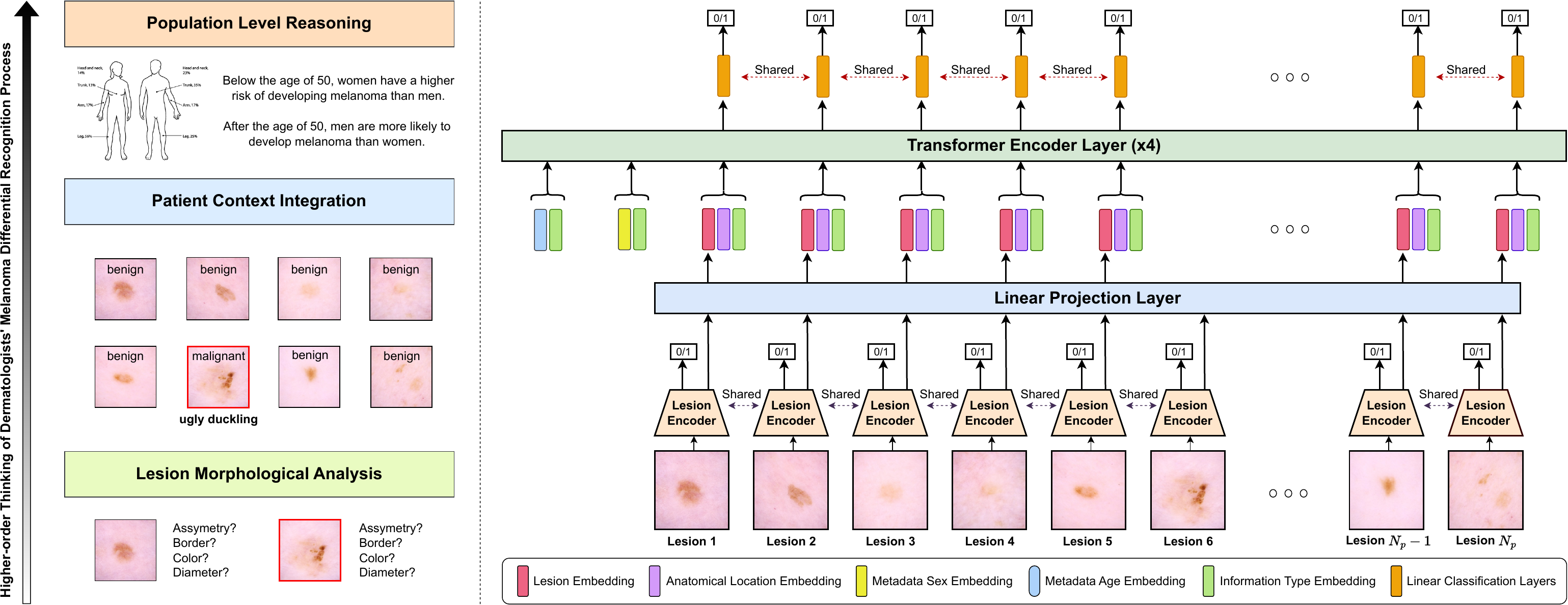}
    \caption{The dermatologists' melanoma diagnostic reasoning process (left) and the pipeline of the proposed MelDD framework (right) inspired by the clinical process. Lesion features are extracted and grouped by patients first, and lesion anatomical site information and patient metadata are incorporated later for enhanced context.}
    \label{fig:your_figure}
\end{figure}

\noindent
\textbf{Modelling the Context of Patient's Skin Ecosystem:} As per the ugly duckling criteria, in a patient, nevi which stand out from the rest are suspicious regardless of morphology. Conversely, lesions which are considered atypical in the absence of patient context may turn out to be normal within the context of a patient. Hence, contextual information is critical. Transformers have demonstrated a remarkable ability to analyze the global context in text, images and videos \cite{neurips_transformer}. Stacked self-attention layers to model dense relations among input tokens allow transformers to capture context information at a patient level. This modelling is used in our design to capture ugly ducklings, if any. We denote the set of extracted (from a CNN) lesion features for a given patient $p$ by the set  $L_p = \{l_{i}^p\}_{i=1}^{N_p}$, where each $l_{i}^p \in L_p$ is obtained by passing the set $X_p$ of lesions from the same patient through a ResNet101 finetuned on SIIM-ISIC 2020 dataset. The dimensionality of each lesion feature is projected onto the dimension $D$ of the transformer. Since the number of lesions per patient is variable, we employ masked self-attention with key padding \cite{neurips_transformer} that applies padding to patients with fewer lesions to align them with $N_{max}^p$, the maximum number of lesions for any patient in the dataset and ignores padding tokens during processing.\\

\noindent
\textbf{Anatomical Site and Masked Self Attention:} Different regions of the body exhibit varying levels of melanoma risk. Hence, the anatomical locations of a lesion can help in ruling out benign lesions. Self-attention \cite{neurips_transformer} generates an attention map of the context utilizing all of the patient's lesions. We use this attention map to implicitly infer the presence of an ugly duckling. To further enhance contextual analysis, we introduce a learnable anatomic site matrix, denoted as $E^L \in \mathbb{R}^{D \times 7}$, which represents the general anatomic sites in our dataset: head/neck, palm/soles, oral/genital, lower extremity, upper extremity, torso, and an additional category for unknown locations. Let  $A_p = \{a_{i}^p\}_{i=1}^{N_p} \in \mathbb{R}^{D \times N_p}$ be the anatomic site representation for patient $p$, obtained by retrieving the corresponding anatomic site embedding for each lesion from $E^L$.The lesion embedding $L_p$ is added to the anatomic site embedding $A_p$, element-wise, to derive an enhanced contextual embedding $Q_p \in \mathbb{R}^{D \times N_p}$: $Q_p = L_p + A_p$. Embedding $Q_p$ is utilized in masked self-attention, generating an attention map that captures \textit{spatial, inter-lesion interactions} within the context of a patient. This integration enables the model to effectively learn the relationship between the anatomical context and individual lesion characteristics at the patient level.\\

\noindent
\textbf{Combining Patient Demographics for Differentials: } Age and sex are risk factors for melanoma, as women have a higher incidence of diagnosis before the age of 50, while men have a higher rate after the age of 50. The incidence of melanoma increases progressively with advancing age, indicating a greater prevalence of melanoma development among individuals as they age \cite{url_stats}. Patients' sex and age information is generally part of the metadata. A learnable embedding $E^S \in \mathbb{R}^{D \times 2}$ is used to represent the male and female sexes. The transformer's trainable embeddings effectively capture and encode the dataset statistics. Positional encodings, incorporating sine and cosine functions, are employed to denote the patient's age through integer binning. The age, sex, and lesions are represented by learning three type embedding vectors, forming the trainable embedding matrix $E^M \in \mathbb{R}^{D \times 3}$, to distinguish one piece of information from another. These distinct type embedding vectors are then added element-wise to the corresponding age, sex, and contextual lesion embeddings, denoted as $S_p$, $Y_p$, and $Q_p$.  The modified embeddings are then concatenated, capturing the combined information of age, sex, and lesions as input to subsequent stages.\\

\noindent
\textbf{Transformer Encoder for Melanoma Recognition: } A multi-layer transformer encoder \cite{neurips_transformer} is composed of a stack of encoder layers, each comprising multi-head self-attention, layer normalization (LN) and feed-forward neural networks (FFN). In the proposed framework, the combined patient representation $E^p = [S_p;  Y_p; Q_p]$ undergoes encoding using a multi-layer transformer encoder. Given input patient representation $E_{l-1}^p$ at the $l^{th}$ layer, \[Encoder(E_{l-1}^p) = E_{l}^p = FFN(LN(Attention(E_{l-1}^p))) + E_{l-1}^p,\] \[Attention = Softmax \left( \frac{E^{pQ}_{l-1} . E^{pK}_{l-1}}{\sqrt{D}} \right) E^{pV}_{l-1}.\]  The contextualized representation of the lesions $E_L^p$ is sent to shared linear layers to perform melanoma recognition. 
\section{Experiments}

\subsection{Data}

The 2020 SIIM-ISIC melanoma recognition dataset \cite{scidata_isic2020} was used for all our experiments. It includes 2,056 patients, among whom 428 individuals exhibit at least one melanoma, with an average of 1.36 melanomas per patient. The dataset comprises 33,126 dermoscopic images, including 584 histopathologically confirmed melanomas, as well as benign lesions that are considered melanoma mimickers such as nevi, atypical melanocytic proliferation, café-au-lait macule, lentigo NOS, lentigo simplex, solar lentigo, lichenoid keratosis, and seborrheic keratosis. Hence, the dataset is severely imbalanced, with melanomas accounting for only $1.8\%$ of the samples.  In addition to the image data, the dataset provides metadata pertaining to the approximate age of patients at the time of capture, their biological sex, and the general anatomical site of the lesion.

\subsection{Experimental Settings}

The dermoscopic skin lesion images were cropped to the maximum square from the centre and resized to 256 $\times$ 256. Our experimental setup involved a patient group stratified five-fold cross-validation without age and sex stratification. Each fold included a designated testing set, while the remaining data was split into $80\%$ for training and $20\%$ for validation. The evaluation on the challenge leaderboard is not conducted due to the unavailability of ground truth for the challenge test set, preventing analysis on our evaluation metrics. The ResNet101 \cite{cvpr_resnet} backbone pre-trained on SIIM-ISIC 2020 dataset to predict lesion-focused recognition was employed for transformer feature extraction. The transformer consisted of 4 layers with 4 MHSA heads, and the model dimension was set to $D$ = 64. The training process utilized the Adam optimizer \cite{arxiv_adam} with a learning rate of $8\mathrm{e}{-5}$, implemented in PyTorch \cite{neurips_pytorch}. It employed the weighted binary cross-entropy loss, based on the inverse of proportions, and was conducted on a single NVIDIA GeForce RTX-2080 Ti GPU. The training, incorporating early stopping, was limited to a maximum of 200 epochs with a batch size of 32.\\

\noindent
\textbf{Metrics:} Many state-of-the-art models for SIIM-ISIC 2020 classification focus on optimizing the area under the ROC curve (AUC). However, this may be inappropriate since AUC is not clinically interpretable \cite{eurrad_rocauc}. For instance, a recent work \cite{tmi_cinet} reports a high AUC score but exhibits poor sensitivity, making it unsuitable for clinical use in melanoma recognition. Additionally, different methods can possess identical AUC values yet perform differently at clinically significant thresholds. To address these limitations, we opt to optimize the balanced accuracy (BACC) at the Youden's J index \cite{cancer_youdens}. This may be more clinically meaningful for a small and imbalanced dataset with low melanoma prevalence $(1.8\%)$ such as SIIM-ISIC 2020 dataset. The operating point determines the cut-off value that minimizes the difference between sensitivity and specificity, better evaluating the clinical utility of diagnostic tests in melanoma recognition.
\section{Results and Discussion}

\begin{table*}[t]
\centering
\caption{Comparision of classification performance in melanoma recognition averaged across five-folds on SIIM-ISIC 2020 dataset: BACC: balanced accuracy, SN: sensitivity, SP: specificity at Youden's J statistic cut-off, and ROC AUC. (PC = patient context, VC = varying lesion count, L = anatomical location, M = metadata).}\label{tab1}
\begin{tabularx}{\textwidth}{hssssbbbb}
\hline
\toprule
MelDD variants  & PC & VC & L & M & BACC    & SN   & SP & AUC  \\
\hline
\midrule
V0 (Baseline) &  \xmark & -- & \xmark & \xmark  &  $0.7649$  & $\textbf{0.8867}$  & $0.6431$  & $0.8371$  \\
V1   &  \cmark & \cmark & \xmark & \xmark &  $0.7841$   &  $0.8679$    & $0.7003$   & $0.8558$  \\
V2  &  \cmark & \cmark & \cmark & \xmark &  $\textbf{0.7904}$  &  $0.8274$    & $\textbf{0.7534}$  & $\textbf{0.8612}$ \\
V3  &  \cmark & \cmark & \xmark & \cmark &  $0.7867$  &  $0.8843$   & $0.6890$   & $0.8544$  \\
V4    &  \cmark & \cmark & \cmark & \cmark &  $0.7793$   &  $0.8761$    & $0.6825$  & $0.8504$  \\
\hline
CI-Net \cite{tmi_cinet}  &  \xmark & -- & \xmark & \xmark & $0.6200$ & $0.3220$   & $0.9180$  & $0.9230$ \\
UDTR-L \cite{miccai_udtr}  &  \cmark & \xmark & \xmark & \xmark &  $0.7564$  & $0.7522$  & $0.7605$  & $0.8493$ \\
UDTR-Adapted  &  \cmark & \xmark & \xmark & \xmark &  $0.7094$   &  $0.7922$    & $0.6266$   & $0.7634$  \\
UDTR-Full \cite{miccai_udtr} &  \cmark & \xmark & \xmark & \xmark &  $0.8183$  &  $0.8164$   & $0.8202$  & $0.8964$  \\
\hline
\bottomrule
\end{tabularx}
\label{results}
\end{table*}

\begin{table*}[t]
\centering
\caption{Performance improvement of the variants over the baseline (in percentage). }\label{tab2}
\begin{tabularx}{\textwidth}{hssssbbbb}
\hline

\toprule
MelDD variants  & PC & VC & L & M & BACC    & SN   & SP & AUC  \\
\hline
\midrule
V0 &  \xmark & -- & \xmark & \xmark  &  0.7649  & 0.8867  & 0.6431  & 0.8371  \\
V1   &  \cmark & \cmark & \xmark & \xmark & $+2.51\%$                      & $-2.12\%$ & $+8.89\%$  & $+2.23\%$ \\
V2  &  \cmark & \cmark & \cmark & \xmark & $\textbf{+3.33\%}$ & $-6.69\%$  & $\textbf{+17.15\%}$ & $\textbf{+2.88\%}$ \\
V3  &  \cmark & \cmark & \xmark & \cmark  & $+2.85\%$                      & $-0.27\%$  & $+7.14\%$  & $+2.07\%$ \\
V4    &  \cmark & \cmark & \cmark & \cmark & $+1.88\%$                      & $-1.20\%$  & $+6.13\%$  & $+1.59\%$ \\
\hline
\bottomrule
\end{tabularx}
\label{results_improvement}
\end{table*}

\begin{figure}[]
    \centering    
    \includegraphics[width=\textwidth]{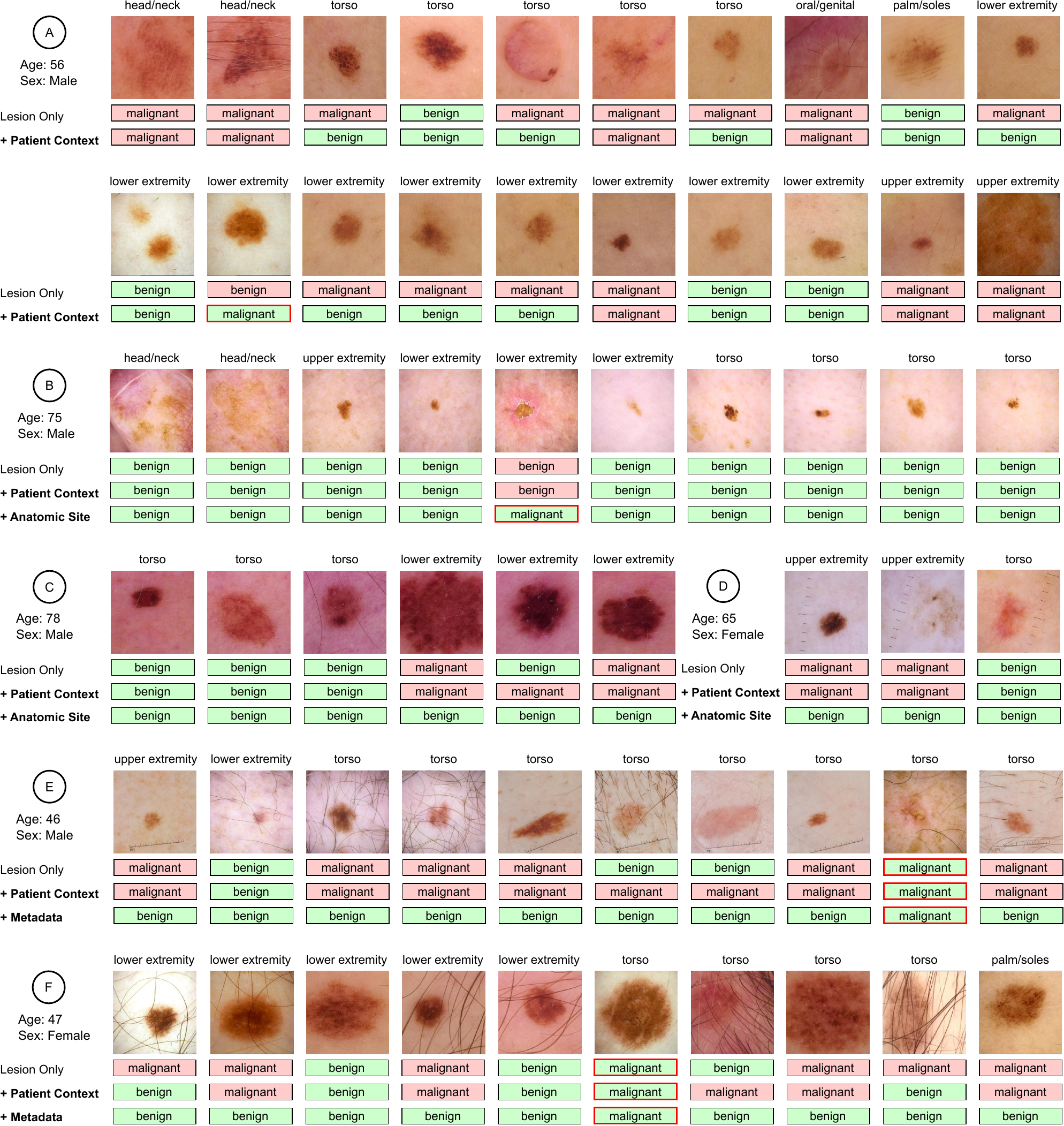}
    \caption{Examples of malignant melanoma prediction changes with additional context and evidence information. Green/red boxes indicate correct/incorrect predictions, respectively. In Patient A, multiple atypical lesions reduce suspicion of malignancy in an additional atypical lesion, while a morphologically typical lesion distinct in the nevus landscape is considered suspicious. Patient B demonstrates how including anatomical location accurately detects an ugly duckling suspicious lesion by comparing it to other lesions in the same location to predict malignancy effectively. The examples of Patients C and D underscore how incorporating location information prevents the misclassification of benign lesions as malignant by considering the specific anatomical characteristics that differentiate suspicious lesions in different locations. Lastly, Patients E and F emphasize the importance of patient demographics to help the model correlate lesion characteristics with susceptibility to risk factors, avoiding misdiagnosis of benign lesions as malignant based on a better understanding of patient-specific factors.}
    \label{fig:qualitative}
\end{figure}

We assess the contributions of the additional information in melanoma recognition using variants of our proposed MelDD framework. These results are presented in Table \ref{results}. Variant V0 (baseline) which solely considers the lesion has a BACC of $76.49\%$ and AUC of $83.71$. This, however, is at a low specificity (SP) of $64.31\%$. Overall, from the figures in Table \ref{results} and \ref{results_improvement}, it can be seen that the addition of information is beneficial as there is a consistent boost in all performance metrics except SN, relative to the baseline. This boost ranges from a modest $1.9\%$ (in BACC for V4) to a significant $17.15\%$ (in SP for V2). The degradation in SN ranges from $0.27\%$ (for V3) to $6.7\%$ (for V2). Including all (lesion, its context, location and metadata) information serves to boost the performance (of V4) by a minimum of $1.6\%$ (AUC) and a maximum of $6\%$ (SP) with a decrease in SN by less than $2\%$.  Figure \ref{fig:qualitative} illustrates patient case studies, demonstrating the impact of incorporating additional context and metadata on melanoma recognition.\\

The obtained results provide sufficient insights that can help in deciding which information is preferable for a specific use case. For instance, the combined knowledge of lesion source (which patient), its characteristics vis a vis other lesions of the patient (to help identify the ugly duckling) and where in the body it is located appears to be best for melanoma diagnosis with an optimal detection threshold, as seen in the figures for MelDD-V2. While balancing both SN and SP is crucial to ensure effective and reliable melanoma diagnosis, their relative importance varies based on priorities. A high SP value will be required to avoid overdiagnosis and needless biopsies. MelDD-V2 is a good choice to meet this requirement. If on the other hand, the application scenario is screening, a higher SN is preferable, and hence, simply using metadata instead of lesion location may be preferable as MelDD-V3 has a high SN and a marginally lower BACC and AUC. This suggests that patient sex and age do play a key role in improving SN. Intuitively, combining all information should be beneficial to performance which is not seen in the result in Table \ref{results}. When we examined the reason for this, we found that there was a sex-wise skew in the melanoma cases in the dataset. A sex-wise stratification in the data split for training/testing could be explored in the future to mitigate the effect of skew.

Finally, we compare the proposed method with the state-of-the-art (SOTA) frameworks, CI-Net \cite{tmi_cinet}, and UDTR \cite{miccai_udtr}. There are some differences in the settings which may impact the comparison. For a start, the SOTA models utilize higher-resolution images compared to our work.  UDTR is designed for a fixed number of lesions; it handles deviation in input through repeated sampling and uses contrastive learning and test-time augmentation techniques. However, repeated sampling in a transformer-based model can lead to overfitting, limited generalization, potential information loss, and difficulties in capturing complete patient context due to random selection and discarding of lesion instances. To ensure fairness, we introduce UDTR-Adapted as a baseline that aligns with our MedDD-V1 while considering a fixed number of lesions. Notably, MelDD-V1 outperforms (in terms of BACC) UTDR-L by $3.66\%$ and  UDTR-Adapted by $10.53\%$ (see the lower part of Table \ref{results}). This highlights the significance of considering the complete patient context.
\section{Conclusion}

Inspired by the clinical diagnostic reasoning process where multiple sources of information are used for diagnosis, we present a modular, multi-level framework for differential diagnosis of malignant melanoma that integrates information at lesion, patient, and population levels. Since the number of lesions a patient may have is unknown, the proposed solution employs a masked transformer to seamlessly incorporate variable lesion counts, enabling flexible integration of patient context information in the decision-making process. Results show the differential roles played by additional information: the context and location information leads to a significant improvement in SP values with a marginal dip in SN, whereas metadata serves to restore SN value to that of the baseline model with a modest increase in SP value. Our results demonstrate that optimising BACC at Youden's J index aids in gaining good control over SP and SN variations. This is in contrast to the conventional approach of optimising AUC, which typically leads to a big tradeoff between SP and SN. Our solution offers a transparent decision support system for melanoma recognition, supporting clinicians in evidence-based decision-making.


\end{document}